# U-DeepONet: U-Net Enhanced Deep Operator Network for Geologic Carbon Sequestration

Waleed Diab[a], Mohammed Al Kobaisi [a,*]

[a] *Petroleum Engineering Department, Khalifa University of Science and Technology, Abu Dhabi 127788, UAE*

*Abstract*

Fourier Neural Operator (FNO) and Deep Operator Network (DeepONet) are by far the most popular neural operator learning algorithms. FNO seems to enjoy an edge in popularity due to its ease of use, especially with high dimensional data. However, a lesser-acknowledged feature of DeepONet is its modularity. This feature allows the user the flexibility of choosing the kind of neural network to be used in the trunk and/or branch of the DeepONet. This is beneficial because it has been shown many times that different types of problems require different kinds of network architectures for effective learning. In this work, we will take advantage of this feature by carefully designing a more efficient neural operator based on the DeepONet architecture. We introduce U-Net enhanced DeepONet (U-DeepONet) for learning the solution operator of highly complex $CO_2$-water two-phase flow in heterogeneous porous media. The U-DeepONet is more accurate in predicting gas saturation and pressure buildup than the state-of-the-art U-Net based Fourier Neural Operator (U-FNO) and the Fourier-enhanced Multiple-Input Operator (Fourier-MIONet) trained on the same dataset. In addition, the proposed U-DeepONet is significantly more efficient in training times than both the U-FNO (more than 18 times faster) and the Fourier-MIONet (more than 5 times faster), while consuming less computational resources. We also show that the U-DeepONet is more data efficient and better at generalization than both the U-FNO and the Fourier-MIONet.

## 1. Introduction

Geological $CO_2$ storage (GCS) stands out as a promising solution to address the accumulation of anthropogenic carbon dioxide in our atmosphere (Bachu, 2008; Benson & Cole, 2008). This method involves the direct injection of $CO_2$ into suitable deep underground geological formations. Prime reservoirs for storage include deep saline aquifers (Pruess & García, 2002), depleted oil and gas fields, and un-mineable coal seams, each selected based on their capacity, injectivity, and long-term retention attributes. As time progresses, the stored $CO_2$ undergoes a sequence of trapping mechanisms, ranging from structural and residual trapping to solubility and eventual mineral trapping, thus enhancing the security of the storage (Saadatpoor et al., 2010). To ensure the integrity and stability of these storage sites, rigorous monitoring and verification protocols are imperative.

Rigorous monitoring and verification protocols play an indispensable role in the efficacy and security of geological $CO_2$ storage operations (Lengler et al., 2010; Strandli et al., 2014). While direct observation methods, such as seismic surveys (Yin et al., 2022) and wellbore monitoring (Fawad & Mondol, 2021), provide valuable real-time data on $CO_2$ plume dynamics and caprock integrity, simulation techniques offer a forward-looking approach to understanding subsurface behavior (Ajayi et al., 2019). Advanced numerical simulations, grounded in reservoir engineering, enable the modeling of $CO_2$ migration, dissolution, and mineralization over time (Flemisch et al., 2023; Tariq, Ali, et al., 2023; Zhao et al., 2023). These simulations, which integrate reservoir characteristics, injection rates, local geology, and many other variables aid in predicting potential leakage paths, pressure build-ups, and interactions between $CO_2$ and formation water. By coupling real-world monitoring data with simulation results, stakeholders can achieve a more holistic understanding of the reservoir storage performance (Anyosa et al., 2021). Such a synergistic approach not only ensures the long-term stability of stored $CO_2$ but also helps in promptly addressing unforeseen challenges.

---

\* Corresponding author at: PE Department, Khalifa University of Science and Technology, Abu Dhabi 127788, UAE
  *E-mail address*: mohammed.alkobaisi@ku.ac.ae (M. Al Kobaisi)

Numerical simulation of geological $CO_2$ storage is a resource-intensive task, demanding both robust algorithms and powerful hardware capabilities. A primary challenge arises from geological uncertainty (Cao et al., 2020; Gan et al., 2021; Jeong et al., 2013; Nordbotten et al., 2012; Xiao et al., 2023). Given that the subsurface cannot be observed directly in detail, our understanding of it is based on sparse observations, which lead to significant uncertainties in reservoir properties like permeability, porosity, residual saturations, etc. To account for these uncertainties, multiple realizations of the geological model are generated, each representing a possible scenario of the subsurface (Mahjour & Faroughi, 2023). However, every realization entails solving coupled partial differential equations that govern fluid flow and transport, and geochemical reactions in porous media. As the number of realizations increases to adequately sample the uncertainty space, the computational expense escalates, often exponentially (Mahjour & Faroughi, 2023). High-resolution models, necessary for capturing fine-scale geological features and processes, further compound the computational demands. Yet, simulating the $CO_2$ injection and migration process for each of these realizations is essential to assess the range of potential outcomes and risks.

Over the past five decades, significant leaps in numerical reservoir simulation have been made. From discretization schemes, and gridding techniques, to solvers, each has grown in sophistication and ability to capture the multi-physics, multi-scale, and nonlinear nature of fluid flow behaviors and interactions in highly heterogeneous porous media (Pruess & García, 2002; Wen et al., 2023). With all these advances, numerical reservoir simulation has become an indispensable tool allowing engineers to simulate and evaluate flow behaviors and patterns. However, as these models increase in sophistication, they become more and more computationally expensive to run. In practice, and as more subsurface data becomes available, a large number of forward simulations (realizations) need to be evaluated.

Powerful central processing units, computer clusters, and parallel computing offer a way to manage the computational demands of simulation (Lichtner et al., n.d.; Zhang et al., 2008). However, even when utilizing state-of-the-art high-performance computing platforms, the integration of fresh data into existing models introduces significant complexities. As more observational data become available—whether from well logs, seismic surveys, or other sources—it necessitates continuous updates to the geological model to refine its accuracy. Yet, assimilating this new information and recalibrating the multitude of realizations to reflect updated knowledge can be computationally intensive, leading to extended simulation times. This iterative model refinement and the associated computational overhead can pose challenges for timely decision-making, especially in scenarios where swift model updates are crucial for operational or safety adjustments (Nordbotten et al., 2012).

In addressing these challenges, researchers are exploring various strategies, such as machine learning-based surrogates and reduced order models (Cardoso et al., 2009; Falola et al., 2023; Ju et al., 2023; Lyu et al., 2023; Stepien et al., 2023; Tang et al., 2022; Tariq, Yan, et al., 2023; Yan et al., 2022; Zhang et al., 2022) to achieve a balance between computational feasibility and simulation fidelity. Nonetheless, as the industry advances towards more extensive and deeper $CO_2$ storage projects, ongoing efforts to optimize and innovate in the realm of computational methods remain paramount. Surrogate models have gained popularity as a way to reduce the computational burden associated with every realization. In essence, surrogate models need to be able to capture as much of the physics of the problem as possible, while maintaining high computational efficiency. The problem is that one is usually promoted at the expense of the other. Neural operator learning algorithms have the potential to solve this problem.

Data-driven neural operator learning algorithms learn the physics of the problem in an implicit manner, i.e. the physics (PDE and coefficients) do not have to be explicitly 'fed into' the model as it can learn them from the data (Alpak et al., 2023; Espeholt et al., 2022; Kissas et al., 2022; Li et al., 2021; Lu et al., 2021a, 2021b; Tripura & Chakraborty, 2022; Wang et al., 2022). This makes data-driven neural operator learning algorithms easier to work with. Physics-informed neural operator learning algorithms, on the other hand, require explicit knowledge of the physics of the problem (Goswami et al., 2022; Wang et al., 2021; Wang & Perdikaris, 2021). As such, the need to generate a sufficiently informative dataset from a sophisticated reservoir simulator is waived, or substantially reduced at the least, for physics-informed algorithms. The main challenge for Data-driven neural operator learning algorithms is generalization given as small a dataset as possible, and for the architecture to be sufficiently efficient so that it can be trained in the shortest possible time with minimum requirements on computer hardware, while maintaining accuracy.

The U-Net Fourier Neural Operator (U-FNO) was recently proposed as an algorithm for operator learning of $CO_2$ geological storage (Wen et al., 2022). The proposed U-FNO was shown to generalize well over 10 input variables for gas saturation and pressure buildup. U-FNO architecture adds an additional U-Net in each Fourier layer which allows the model to process the inputs more effectively. The effectiveness of the U-Net stems from its ability to process data on structured grids through local convolution, and hence enriching the representation power of the architecture in higher frequencies leading to better accuracy compared to FNO and convolutional-FNO (conv-FNO) (Wen et al., 2022). However, the proposed U-FNO has three main drawbacks; 1. An extremely large number of trainable parameters (more than 30 million in this case), 2. Difficulty in scaling to higher dimensions, and 3. Loss of the resolution invariance of FNO in time, i.e., it cannot make predictions at unseen time steps. These three drawbacks are mainly due to the architectural choices made when designing the U-FNO.

In a recent work, Jiang et al. (2023) proposed to combine the powerful U-FNO with the multiple-input deep neural operator (MIONet) of Jin et al. (2022), and the authors termed their architecture Fourier-enhanced multiple-input neural operator (Fourier-MIONet). The introduction of the MIONet to the U-FNO architecture to create the Fourier-MIONet is a more computationally and data-efficient alternative to the U-FNO alone, albeit it results in a small loss of accuracy. In this work, we propose U-DeepONet, a U-Net enhanced DeepONet that offers significant gains in computational efficiency and accuracy over both the U-FNO and Fourier-MIONet.

The paper is organized as follows. In Section 2, we introduce the mathematics of flow and transport of $CO_2$ for geological storage and the dataset generated using numerical simulation to train and evaluate the various models. In Section 3, we review the U-FNO and the Fourier-MIONet, and then we introduce our U-DeepONet. In Section 4, we show the efficiency contrast between the three models and we report on the predictive accuracy; additional benefits of the U-DeepONet are discussed in Section 5 and we conclude the paper in Section 6.

## 2. Problem Setup

### 2.1. Governing Equations

In geological storage of $CO_2$, predicting the migration, trapping, and long-term fate of injected $CO_2$ is of paramount importance. This is often achieved through numerical simulations by solving mass conservation equations, considering the interactions between phases, the porous medium, and various physical effects like capillarity, dissolution, and buoyancy. The mass accumulation equations for a multi-component flow problem involving $CO_2$ and formation water can be written for each component $\eta$ as (Zhang et al., 2008):

$$\frac{\partial(\emptyset \sum_p S_p \rho_p X_p^\eta)}{\partial t} = -\nabla \cdot \left(\mathbf{F}^\eta|_{adv} + \mathbf{F}^\eta|_{dif}\right) + q^\eta, \tag{1}$$

where $\mathbf{F}^\eta|_{adv}$ is the advective mass flux, $\mathbf{F}^\eta|_{dif}$ is the diffusive mass flux, $q^\eta$ is the volumetric flow rate of the injection source, $\emptyset$ is the porosity; for each wetting phase $p$: $S_p$ is the saturation, $\rho_p$ is the density, and $X_p^\eta$ is the mass fraction of component $\eta$ in phase $p$.

For simplicity, Wen et al. (2022) opted not to explicitly include molecular diffusion and hydrodynamic dispersion in the simulation, thus $\mathbf{F}^\eta|_{dif} = 0$. However, some diffusion and dispersion effects still exist due to numerical artifacts produced by the finite difference scheme used in the spatial discretization of the system of equations. The advective mass flux for each component $\mathbf{F}^\eta|_{adv}$ is obtained by summing over the phases $p$,

$$\mathbf{F}^\eta|_{adv} = \sum_p X^\eta \mathbf{F}_p = \sum_p X^\eta \left(-k \frac{k_{r,p}}{\mu_p} \rho_p (\nabla P_p - \rho_p g)\right). \quad (2)$$

Here, $\mathbf{F}_p$ is the multiphase Darcy's law for each phase $p$, where $k$ is the absolute permeability; for each phase $p$: $k_r$ is the relative permeability which is a non-linear function of $S$, $\mu$ is the viscosity which is a function of $P$, and $g$ is the gravitational acceleration. Pressure ($P_p$) and saturation ($S_p$) are coupled through the capillary pressure $P_c$, which is defined for wetting ($w$) and non-wetting ($n$) phases as follows:

$$P_c(S_p) = P_n - P_w, \quad (3)$$

where $P_c$ is a non-linear function of $S_p$. To compound the non-linearity in this problem, porosity $\emptyset$, density $\rho$, and $CO_2$ solubility are also non-linear functions of pressure $P_p$.

## 2.2. Dataset

We use the open-source dataset published by Wen et al. (2022). The dataset, which contains 5500 realizations of $CO_2$ geological storage in a radial and symmetric reservoir, is generated using the numerical reservoir simulator ECLIPSE (e300) (Schlumberger, 2014). The purpose of these realizations is to track the movement of the $CO_2$ plume and pressure buildup over time. The authors used two hundred gradually coarsened grid cells in the radial direction with a 100,000 m radius, and they solve the system at 24 gradually coarsening time snapshots $\{1\ day, 2\ days, 4\ days, \dots, 14.8\ years, 21.1\ years, 30\ years\}$ using an adaptive implicit method for temporal discretization. Supercritical $CO_2$ is injected at a constant rate ranging from 0.2 to 2 Mt/year for a period of 30 years. The super critical $CO_2$ is injected through a vertical well with a radius of 0.1 m. The realizations account for a variable perforation thickness up to the entire thickness of the reservoir which can also vary from 12.5 to 200 m. The outer boundaries of the reservoir are closed (no-flow boundaries).

The dataset contains 5500 input-to-output mappings for saturation and another 5500 input-to-output mappings for pressure buildup; each dataset is split as 9:1:1, where 4500 realizations are used for training, 500 realizations for validation, and 500 realizations for testing. The outputs are the state variables: gas saturation ($S_g$) and pressure buildup ($dP$), the inputs consist of 9 variables: four field variables (space-dependent inputs) and five scalar variables (Table 1). In theory, the learned operator should be able to generalize over these nine inputs; in other words, given a new combination of these nine variables that do not exist in the training dataset the learned operator should accurately predict the state variables. The field variables include a horizontal permeability map ($k_x$), a vertical permeability map ($k_y$), a porosity map ($\emptyset$), and an injection perforation map ($perf$). Moreover, scaler variables include the initial reservoir pressure at the top of the reservoir ($P_{init}$), injection rate ($Q$), reservoir temperature ($T$), capillary pressure scaling factor ($\lambda$), and irreducible water saturation ($S_{wi}$). We direct the readers to the original paper (Wen et al., 2022) for more details on the generation of the field maps and all other sampling techniques for the inputs. An example of an input-to-output mapping is shown in Fig. 1.

**Table 1:** Summary of the network input variables. Field inputs are maps on a (96, 200) grid. $\mathcal{V}[a,b]$ means that the value ranges from a to b. Note that while the perforation height and location are sampled as scalars, they are represented as a binary map and therefore we count them as the forth field input.

|  | Parameter | Notation | Distribution | Unit |
|---|---|---|---|---|
| Field | Horizontal permeability field | $k_x$ | $\mathcal{V}[0.001, 10000]$ | $mD$ |
|  | Material anisotropy ratio | $k_x/k_y$ | $\mathcal{V}[1, 150]$ | — |
|  | Porosity | $\emptyset$ | — | — |
| Scalar | Injection rate | $Q$ | $\mathcal{V}[0.2, 2]$ | $MT/y$ |
|  | Initial pressure | $P_{init}$ | $\mathcal{V}[100, 300]$ | $bar$ |
|  | Iso-thermal reservoir temperature | $T$ | $\mathcal{V}[35, 170]$ | °C |
|  | Irreducible water saturation | $S_{wi}$ | $\mathcal{V}[0.1, 0.3]$ | — |
|  | Van Genuchten scaling factor | $\lambda$ | $\mathcal{V}[0.3, 0.7]$ | — |
|  | Perforation top location | $perf_{top}$ | $\mathcal{V}[0, 200]$ | $m$ |
|  | Perforation bottom location | $perf_{bottom}$ | $\mathcal{V}[0, 200]$ | $m$ |

Additionally, reservoir thickness ($b$) is randomly sampled between 12.5 and 200 m, with a vertical grid thickness of 2.08 m; this means that the number of vertical grid cells also varies between 6 and 96 and if the number of vertical grids in a specific case is less than 96, the difference is accounted for via zero padding (mask) as in [43]. Consequently, all realizations are of consistent resolution (96, 200).

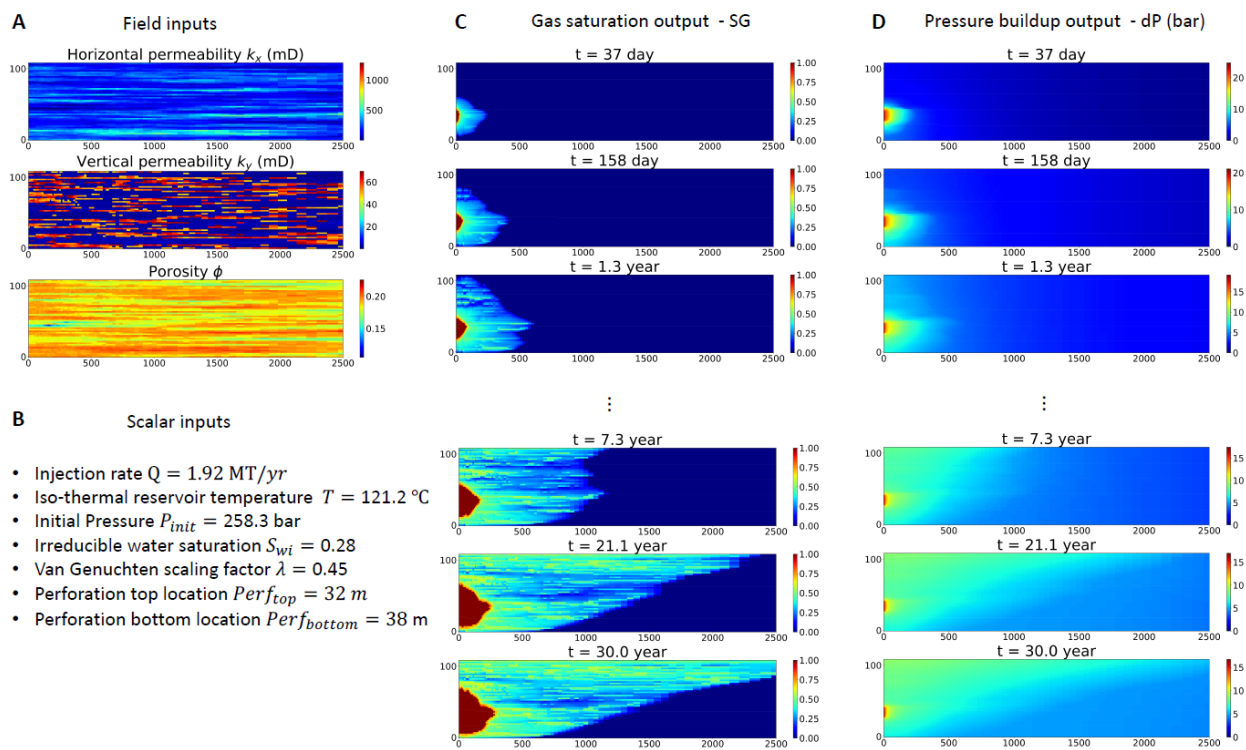

**Figure 1:** An example of an input-to-output mapping. **A.** Field input variables. **B.** Scalar input variables. **C.** Time snapshots of $CO_2$ saturation outputs. **D.** Time snapshots of pressure buildup outputs (Jiang et al., 2023).

## 3. Methods

The U-FNO was recently introduced as a surrogate model for $CO_2$ geological storage. It demonstrated the ability to effectively and accurately generalize to unseen instances of parametric PDEs (Wen et al., 2022). However, the U-FNO suffers from severe computational inefficiencies and its limited ability to generalize to unseen time steps. Building on this, Jiang et al. (2023) integrated the capabilities of U-FNO with the multiple-input deep neural operator (MIONet), leading to the creation of a Fourier-enhanced multiple-input neural operator (Fourier-MIONet). The Fourier-MIONet brings enhanced computational efficiency to the training phase, evidenced by its performance being twice as fast, per epoch, as the U-FNO. This translates to a reduction in training time per epoch from 1535 seconds with U-FNO to 730 seconds on an NVIDIA GeForce RTX 3090 GPU. Additionally, it regains the capacity to generalize over unseen time steps; although these advantages come at a slight cost to accuracy. The training duration reduced from approximately 42-43 hours for the U-FNO to a more manageable 16-17 hours for the Fourier-MIONet on the same GPU. Yet we believe this improvement remains suboptimal.

In this work, we introduce a novel U-DeepONet. By leveraging the inherent modularity of the DeepONet, we incorporate a U-Net within its branch structure. This new approach presents substantial improvements in computational efficiency compared to both the U-FNO and Fourier-MIONet. Furthermore, it retains the accuracy standards set by the U-FNO, while also inheriting the other merits of the Fourier-MIONet. In this section, we first introduce U-FNO and Fourier-MIONet. A more effective and efficient approach for data-driven operator learning, the U-Net enhanced deep operator network (U-DeepONet), is then presented.

### 3.1. Operator Learning

Neural operator learning refers to the process of using deep learning techniques to approximate and learn operators. In functional analysis and differential equations, operators are infinite-dimensional space mappings between finite function spaces. For instance, a differential operator transforms a function ($u$) into its derivative ($v$). Neural operator learning is about training a neural network to emulate or approximate this infinite-dimensional differential operator given a finite set of input-output pairs. It is important to highlight the key difference between operator regression and function regression; the latter only maps one function to another. We refer to this operator as $\mathcal{G}$ and to its approximate (learned) as $\mathcal{G}_\theta$.

We aim to learn (approximate) the nonlinear operator $\mathcal{G}_\theta: \mathcal{U} \to \mathcal{S}$ subject to some loss function, such that

$$\mathcal{G}_\theta(u) = s(u), \qquad (4)$$

where, $u \in \mathcal{U}$ denotes a function in the input functions space, and $\mathcal{G}(u) \in \mathcal{S}$ denotes the corresponding unknown solution in the output space. Both $\mathcal{S}$ and $\mathcal{U}$ are separable Banach spaces of functions that take values in $\mathbb{R}^{d_s}$ and $\mathbb{R}^{d_u}$ respectively. For example, $u$ would be a function, and $s = \mathcal{G}_\theta(u)$ is its derivative. In this paper, $u$ represents the field and scalar variables described in Section 2.2; $s$ represents the spatiotemporal gas saturation and pressure buildup solutions to complete the input-output pairs required to train the operator $\mathcal{G}_\theta$, $u$ is evaluated at $x$ which is a collection of fixed points (grid).

A key advantage of operator learning is that once a model is trained, it can generalize to new input functions. Thus, in inference, a trained operator is orders of magnitude faster than a numerical solver. Another key advantage of operator learning is that it can be trained using simulation data, experimental (real/noisy) data, or both. It can also be informed of the underlaying partial differential equation (PDE) if it is known, which helps it generalize even better.

### 3.2. U-FNO

The U-Net Fourier Neural Operator (U-FNO) was recently proposed by Wen et al. (2022) for operator learning in $CO_2$ geological storage. The U-FNO is an extension of the FNO with an additional U-Net inserted in all or some of the Fourier layers as shown in Fig. 2. The Fourier Neural Operator or FNO was first proposed by Li et al. (2021). The key feature of the FNO is that it formulates the operator by parameterizing the integral kernel directly in Fourier space. This means that the parameters of the network are defined and learned in the Fourier space and the coefficients corresponding to the Fourier series representation of the output function are inferred from the dataset.

The U-FNO architecture applies to the input iteratively (see Fig. 2); it entails the following:

1. Lift the input observations $v(x)$ to a higher dimensional representation $z_{l_0}(x)$ :

$$z_{l_0}(x) = P(v(x)) \in \mathbb{R}^{d_z}, \tag{5}$$

   where $P$ is a linear layer or shallow fully-connected neural network. Here, both $z_0$ and $v$ are defined on the same mesh, and $P$ is a local transformation $P: \mathbb{R} \to \mathbb{R}^{d_z}$; the result of this transformation ($z_0$) can be viewed as an image with $d_z$ channels.

2. Apply $L$ Fourier layers to $z_0$ (Fig. 2B). The output of the $l$th Fourier layer is $z_l$ with $d_v$ channels. Each Fourier layer entails a Fast Fourier Transform (FFT) referred to as $\mathcal{F}$, an inverse FFT $\mathcal{F}^{-1}$, and a weight matrix $\mathcal{R}$ :

$$\mathcal{F}^{-1}(\mathcal{R}_l \cdot \mathcal{F}(z_l)). \tag{6}$$

   Here, $\mathcal{F}$ is applied to each channel of $z_l$, only the first $k$ Fourier modes are used and the remaining are truncated. For 2D inputs $k = k_1 \times k_2$, and for 3D inputs $k = k_1 \times k_2 \times k_3$, where $k_i$ is the number of Fourier modes to be used in each dimension $i$. This means that $\mathcal{F}(z_l)$ has a dimension $d_v \times k$. The weight matrix $\mathcal{R}$ consists of complex-numbers with a shape of $d_v \times d_v$. There are $k$ trainable weight matrices $\mathcal{R}$ forming a weight tensor $\mathcal{R}_l \in \mathbb{C}^{d_v \times d_v \times k}$. This results in $\mathcal{R}_l \cdot \mathcal{F}(z_l)$ and $\mathcal{F}(z_l)$ having the same shape. The tensor $\mathcal{R}_l \cdot \mathcal{F}(z_l)$ is appended with zeros to fill in the truncated modes.

   Each Fourier layer also contains a residual connection $W_l \cdot z_l$ which has the same shape as $z_l$, where $W_l \in \mathbb{R}^{d_v \times d_v}$ is a weight matrix that is used to compute a new set of $d_v$ channels. Each new channel is a linear combination between all the $z_l$ channels. The output of the $(j+1)$th Fourier layer $z_{l_{j+1}}$ is defined as follows:

$$z_{l_{j+1}} = \sigma\left(\mathcal{F}^{-1}\left(\mathcal{R}_{l_j} \cdot \mathcal{F}(z_{l_j})\right) + W_{l_j} \cdot z_{l_j} + b_{l_j}\right), \tag{7}$$

   where $\sigma$ is a nonlinear activation function, and $b_l \in \mathbb{R}^{d_v}$ is a bias.

3. Apply $M$ U-Fourier layers to the output of the last Fourier layer $z_L$ (Fig. 2C). The U-Fourier layer $z_{m_{k+1}}$ is defined based on the Fourier layer and a 3D U-Net as follows:

$$z_{m_{k+1}} = \sigma\left(\mathcal{F}^{-1}\left(\mathcal{R}_{m_k} \cdot \mathcal{F}(z_{m_k})\right) + U_{m_k}(z_{m_k}) + W_{m_k} \cdot z_{m_k} + b_{m_k}\right), \tag{8}$$

   where $U_{m_k}$ is a U-Net block. The second residual connection $U_{m_k}(z_{m_k})$ has the same shape as $z_m$.

4. The output of the last U-Fourier layer $z_M$ is projected back to the original space via a linear transformation parametrized by a linear layer or shallow fully-connected neural network $Q: \mathbb{R}^{d_z} \to \mathbb{R}$ as follows:

$$u(x) = Q(z_M(x)). \tag{9}$$

A few points to note here on the efficiency of FNO and U-FNO. First, it is clear that the Fourier layer is computationally expensive due to the large number of trainable parameters, as there are two weight matrices for every Fourier layer. In addition, the size of the weight tensor $\mathcal{R}_l$ grows exponentially with the increase in the number of Fourier modes ($k_i$), i.e., the Fourier based neural operators suffer from the curse of dimensionality. These issues are only exacerbated with a U-Fourier layer, as there are three trainable weight matrices. Moreover, the U-Net block in the U-Fourier layer could potentially be very expensive to compute if the input data is in 3D as in the case of the U-FNO paper (Wen et al., 2022); 3D convolution is computationally very expensive. These sources of inefficiency contribute to high training times reported for the U-FNO architecture. In this work, we design our architecture with the aforementioned points in mind to avoid these computational bottlenecks.

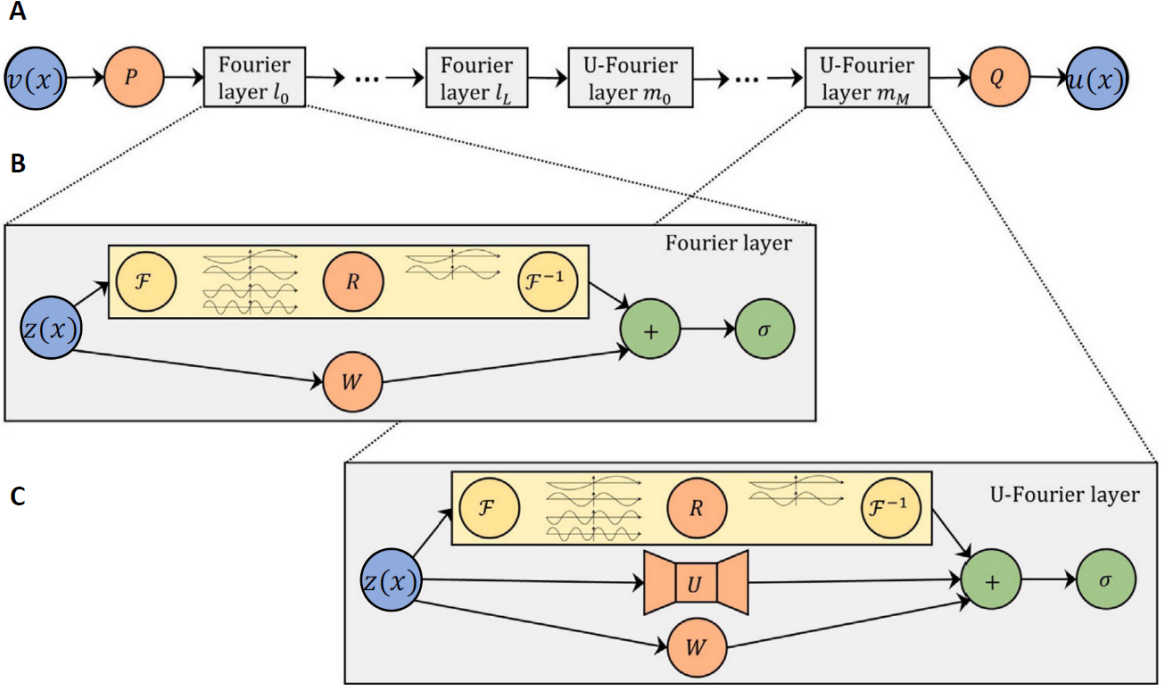

**Figure 2:** Architecture of U-FNO. In (A) the full U-FNO architecture is shown, where v(x) is the input function, P is a fully connected neural network that lifts the input to a higher dimensional space, followed by Fourier layers then U-Fourier layers; Q is a fully connected neural network that maps the output u(x) to the original dimensional space. (B) is the Fourier layer, where $\mathcal{F}$ denotes the Fourier transform, $\mathcal{R}$ is a weight matrix, $\mathcal{F}^{-1}$ is the inverse Fourier transform, W is another weight matrix, and σ is the activation function. (C) is the U-Fourier layer, U denotes the additional U-Net block in the Fourier layer. Figure is adapted from (Wen et al., 2022).

### 3.3. Fourier-MIONet

The Fourier-enhanced Multiple-Input Operator (Fourier-MIONet) architecture design is inspired by Multiple-Input deep Operators (MIONet) to address some of the computational inefficiencies of the U-FNO. Jiang et al. (2023) attached a MIONet to the input of the U-FNO architecture. The MIONet was proposed earlier by Jin et al. (2022) as an extension to the DeepONet to learn multiple nonlinear operators between function spaces. As opposed to the DeepONet which learns to map from a single Banach space, the MIONet was theoretically shown to be able to map between multiple Banach spaces.

#### 3.3.1. Vanilla DeepONet

The stacked DeepONet (Lu et al., 2021a) architecture is simple, it consists of two deep neural networks which output is joined via an inner product. The first neural network, the trunk, processes the coordinates $y \in D'$ of the spatiotemporal domain $y$ as inputs, and learns features embedding $[t_1, t_2, ..., t_q]^T \in \mathbb{R}^q$. The second neural network, the branch, processes the $n$ input functions $v$ discretized as $[v(x_1), v(x_2), ..., v(x_m)]$ at a collection of fixed locations $\{x_i\}_{i=1}^m$ and learns features embedding $[b_1, b_2, ..., b_q]^T \in \mathbb{R}^q$ as output. Each input function $v_i$ is defined on the domain $D \subset \mathbb{R}^{d_i}$. Thus, the input output mapping is defined as follows:

$$u: D' \ni y \rightarrow u(y) \in \mathbb{R}. \tag{10}$$

The joined output $\mathcal{G}_\theta$ of the two networks for the prediction of a function $v$ evaluated at $y$ is defined as:

$$\mathcal{G}_\theta(v)(y) = \sum_{k=1}^{q} b_k\big(v(x_1), v(x_2), \ldots, v(x_m)\big) t_k(y) + b_0, \tag{11}$$

where $b_0 \in \mathbb{R}$ is a bias term.

DeepONet differs from the FNO and U-FNO in two fundamental ways. First, the DeepONet is a high-level framework in which the architecture of the trunk and branch networks is not restricted to any specific kind of neural network. This means that the choice of branch and trunk networks is problem dependent. We capitalize on this property in designing the U-DeepONet. Second, the input domain $D \subset \mathbb{R}^d$ is different from the output domain $D' \subset \mathbb{R}^d$. This means that the output of the DeepONet does not need to be discretized, unlike the FNO. It also means that the DeepONet can make predictions at any location. The Fourier-MIONet architecture takes advantage of this property, as does the U-DeepONet, to lower the computational burden.

### 3.3.2. MIONet

The MIONet (Jin et al., 2022) is an extension of the DeepONet; it consists of several branch networks and a single trunk network. This allows it to learn multiple-input operators $\mathcal{G}$. There are two main changes here, first, the input domain $D_i \subset \mathbb{R}^{d_i}$ is defined for $n$ input functions $v_i$ for $i = \{1, \ldots, n\}$ where:

$$v_i : D_i \to \mathbb{R}. \tag{12}$$

Similarly, the multiple-input operators $\mathcal{G}$ are defined on the product of Banach spaces:

$$\mathcal{G} : X_1 \times X_2 \times \ldots \times X_n \to Y \tag{13}$$

where $X_1, X_2, \ldots \times X_n$ are $n$ different Banach spaces, and $Y$ is the output Banach space.

Second, the input-output function mapping becomes:

$$\mathcal{G} : (v_1, \ldots, v_n) \to u \tag{14}$$

The joined output operator $\mathcal{G}_\theta$ of the $n$ branches and the trunk for the prediction of a function $v$ evaluated at $y$ is defined as:

$$\mathcal{G}_\theta(v)(y) = \sum_{k=1}^{q} b_k^1(v_1) \times b_k^2(v_2) \ldots \times b_k^n(v_n) \times t_k(y) + b_0. \tag{15}$$

Like the DeepONet, the MIONet is a high-level framework in which the architecture of the trunk and branch network is not restricted to any specific kind of neural network. In addition, all the techniques developed for DeepONet can be directly used for MIONet, as well as for our proposed U-DeepONet.

### 3.3.3. Fourier-MIONet

The goal of the Fourier-MIONet architecture is to tackle three main weaknesses of the U-FNO;

1. The computational expense associated with the use of 3D convolution.
2. The computational expense associated with the relatively high number of Fourier modes.
3. The resolution invariance imposed on the time dimension by the U-Net.

Jiang et al. (2023) designed the Fourier-MIONet architecture specifically to tackle these inefficiencies. The Fourier-MIONet architecture utilizes a modified MIONet to process the inputs then passes the output to a U-FNO, creating a hybrid architecture (Fig. 3). This is achieved by first, separating the time variable $t$ from the other input variables and processing it through the trunk network, whose output is $u(\cdot,\cdot,t) \in \mathbb{R}^{96 \times 200}$. In contrast, the output of the U-FNO network would be $u(\cdot,\cdot,\cdot) \in \mathbb{R}^{96 \times 200 \times 24}$, while the MIONet or a vanilla DeepONet output would be $u(x, y, t) \in \mathbb{R}$. By separating the time variable from the other input variables, 2D convolution can be used in the U-Net block instead of 3D convolution. In addition, one of the Fourier modes is now removed ($k = k_1 \times k_2$), which significantly reduces the size of the $\mathcal{R}_l$ tensor. This results in significant improvements in efficiency by dropping the number of trainable parameters from 30+ million to 3+ million with a small loss of accuracy. Second, they use two branch networks $b_1$ and $b_2$ to process the two different kinds of inputs, the field inputs $v_1$ and scalar inputs $v_2$. They opt for a point-wise summation to merge the output of the two branches. The joined output $\mathcal{Z}$ of the 2 branches and the trunk is as follows:

$$\mathcal{Z} = \sum_{k=1}^{q} \left( b_k^1(v_1) + b_k^2(v_2) \right) \times t_k(y) + b_0. \tag{16}$$

To this end, the MIONet is used as an encoder network $\mathcal{Z}$. To map the hidden vector $\mathcal{Z}$ to the output $u$, the U-FNO architecture is applied iteratively as outlined in Section 3.2 with appropriate changes.

It should be noted that the choice to use a point-wise summation to merge the output of the two branches is not covered by the theory of MIONet. The theoretical basis of MIONet was developed for the product of Banach spaces, and the (low-rank) MIONet proposed by Jin et al. (2022) uses a Hadamard product (element-wise product) to merge the branch networks. Despite the lack of theoretical basis, the proposed Fourier-MIONet has been shown to work effectively. An additional advantage of using DeepONet-style input is that the Fourier-MIONet regains the ability of the architecture to interpolate the solution in time $t$.

Jiang et al. (2023) also showed that it would be easy to scale the problem to higher dimensions. For instance, a third dimension in space can be handled by the trunk network to avoid expensive 3D convolution in the U-Net. This feature is also present in our U-DeepONet. In this work, we propose to take full advantage of the DeepONet and to incorporate a U-Net in the branch, we term this architecture U-DeepONet as opposed to the U-FNO. We postulate that it is unnecessary to process the inputs through Fourier layers to achieve similar accuracy and even more significant gains in efficiency.

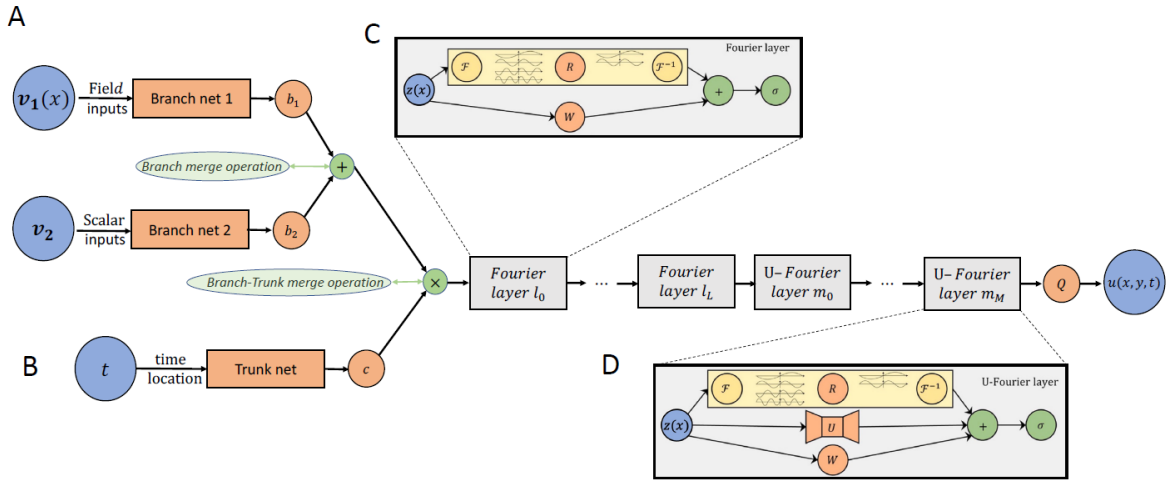

**Figure 3:** Architecture of Fourier-MIONet. **A.** $v_1$ and $v_2$ are the field and scalar inputs, respectively. **B.** Time input $t$. **C.** Fourier layer. **D.** U-Fourier layer. Figure is adapted from (Jiang et al., 2023).

### 3.4. U-DeepONet

Inspired by the DeepONet, the ideas implemented in Fourier-MIONet, and U-FNO architectures, we propose a novel architecture that is based on the DeepONet and the U-Net; we call this architecture the U-DeepONet shown in Fig. 4. We posit that using the Fourier layer and the MIONet is completely unnecessary, and computationally expensive. We build on the idea proposed for Fourier-MIONet of processing the time variable $t$ through the trunk of the DeepONet to leverage the computational benefits of using 2D convolution instead of the 3D convolution. In addition, we leverage multiple U-Net blocks in the branch of the DeepONet to process both the field and scalar inputs. This setup allows for multiple input operator learning without the MIONet architecture. However, the theoretical analysis is left for a future work.

In the U-DeepOnet, the trunk network $t_k(t)$ is a simple feed-forward neural network (FNN) with a nonlinear activation function $\sigma$ and $t$ is the time variable. The branch of the U-DeepONet consists of several blocks connected in series:

1. Lift the input observations $v(x)$ to a higher dimensional representation $b_0(x)$:

$$b_0(x) = P(v(x)) \in \mathbb{R}^{d_z}, \quad (17)$$

   where $P$ is a linear layer, and it is a local transformation $P: \mathbb{R} \to \mathbb{R}^{d_p}$; the result of this transformation ($b_0$) can be viewed as an image with $d_p$ channels. The purpose of this mapping is to increase the number of channels so that a bigger weight tensor can be constructed in the U-Net blocks. In addition, it allows the U-DeepONet to process multiple input operators simultaneously.

2. Apply $L$ U-Nets to $b_0$. The output of the $l$th U-Net block is $U_L$ with $d_z$ channels.

$$U_{l_{j+1}} = \sigma\left(U_{l_j}\right), \quad (18)$$

   where $\sigma$ is a nonlinear activation function. In our experiments we did not observe significant performance changes with different types of activation functions, but the choice of activation functions is always problem specific.

3. The operator $\mathcal{G}_\theta$ is defined as the inner product of the branch and trunk networks:

$$\mathcal{G}_\theta(v)(t) = \sum_{k=1}^{q} U_L(v) t_k(t). \quad (19)$$

4. The output of the product of the trunk and the branch $\mathcal{G}_\theta(v)(t)$ is projected to the solution space via a linear transformation parametrized by a linear layer or shallow fully-connected neural network layer $G: \mathbb{R}^{96 \times 200 \times 24 \times d_z} \to \mathbb{R}^{96 \times 200 \times 24 \times 1}$:

$$u(x, y, t) = G(\mathcal{G}_\theta(v)(t)). \quad (20)$$

A close examination of our novel U-DeepONet reveals that not only does our architecture no longer use 3D convolution in the U-Net block, it also does not make use of the Fourier layer, its associated weight tensors, and FFT. This contributes to significant gains in computational efficiency compared to the U-FNO, and Fourier-MIONet. The basic idea here is that the DeepONet on its own is an effective operator learning framework, and using a combined DeepONet-Fourier based architecture is redundant, as both are neural operator learning networks. In the results section we will demonstrate the effectiveness of the U-DeepONet and the gains in efficiency achieved while maintaining the same order of accuracy as the U-FNO.

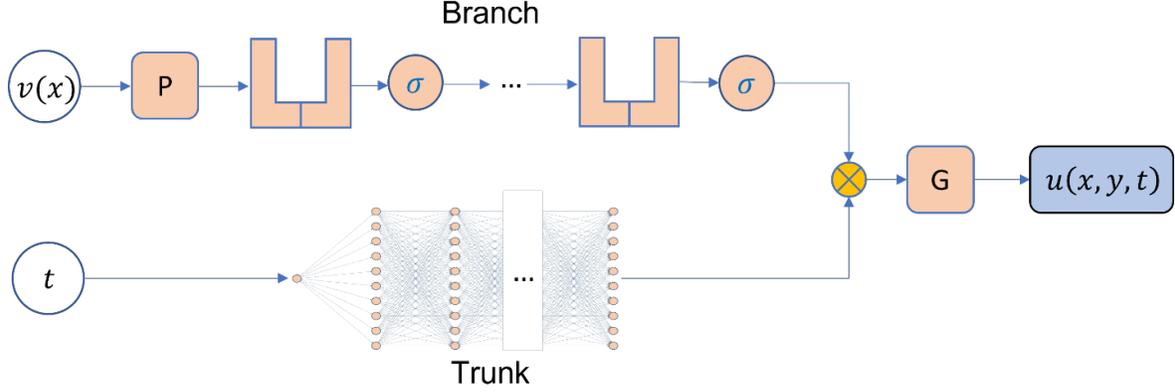

**Figure 4:** U-DeepONet architecture. The figure shows the branch and the trunk networks. The trunk is a feed forward neural network, where $t$ is the time input. In the branch, P is a linear layer, followed by U-Net blocks; $\sigma$ is an activation function, and G is a shallow neural network.

### 3.5. Training and Hyperparameters

In the interest of consistency, we follow the approach presented in (Wen et al., 2022) by constructing a mask to account for the varying thicknesses of the various realizations in the dataset. As such, the loss is only computed within that mask during training for each realization, while the cells laying outside each reservoir are padded with zeros. Hence, the $l_p$-loss is given by:

$$L(y, \hat{y}) = \frac{\|y - \hat{y}\|_p}{\|y\|_p} + \beta \frac{\|dy/dr - d\hat{y}/dr\|_p}{\|dy/dr\|_p}, \tag{21}$$

where $y$ and $dy/dr$ are the ground truth and the first derivative of the ground truth, respectively, $\hat{y}$ and $d\hat{y}/dr$ are the predicted output and the first derivative of the predicted output, respectively, $p$ is the order of the norm, and $\beta$ is a hyperparameter. By choosing $\beta = 2$, we retrieve the relative $L_2$ loss. The first derivative term has been shown to greatly improve the prediction accuracy and promote faster convergence (Wen et al., 2022).

  We train two separate U-DeepONet networks for saturation and pressure buildup. The trunk in both models consists of a feed forward neural network with 10 layers, while the branch consists of three U-Nets connected in series as described in Section 3.4. Moreover, we use two different learning schedules for saturation and pressure buildup and the training is stopped once the loss stops decreasing. To allow for fair comparisons with the U-FNO and Fourier-MIONet, in general we try to avoid changing hyperparameters where possible. Therefore, we maintain the same U-Net structure, number of epochs, and batch size as was reported in (Jiang et al., 2023; Wen et al., 2022). It is worth noting that although the FNO architecture does not require spatial data as input, both Wen et al. (2022) and Jiang et al. (2023) use it as an additional input feature. This has been shown to improve accuracy, which is also consistent with other reports in the literature (Lu et al., 2022). In this work, we also use the spatial data as an additional input feature in the branch for consistency.

### 3.6. Performance Evaluation

To evaluate the performance of each trained model, we use a 9:1:1 data split (Wen et al., 2022) with 500 validation realizations and another 500 test realizations. We benchmark the performance of the trained gas operator networks using the plume area coefficient of determination $R^2_{plume}$, the plume mean absolute error (MPE), and the mean absolute error (MAE). The plume area is defined as non-zero values in either ground truth or prediction (Wen et al., 2022). This approach is used to only evaluate the gas saturation model's accuracy because the gas saturation outside the $CO_2$ plume is always zero. For the evaluation of the pressure buildup model, we use the field mean relative error ($MRE$), $R^2$ score, and the MAE. We follow the definitions in (Tang et al., 2020; Wen et al., 2022) of MRE for pressure buildup:

$$MRE = \frac{1}{n_t n_e n_b} \sum_{t=1}^{n_t} \sum_{i=1}^{n_e} \sum_{j=1}^{n_b} \frac{|\hat{p}_{i,j}{}^t - p_{i,j}^t|}{p_{i,max}^t - p_{i,min}^t}, \quad (22)$$

where $n_t = 24$ is the total number of time steps, $n_e = 500$ is the total number of test samples, $n_b = 96 * 200 = 19{,}200$ is the number of grid blocks, $\hat{p}_{i,j}{}^t$ and $p_{i,j}^t$ denote the pressure value provided by the model and the ground truth, respectively, for test sample $i$, in grid block $j$, at time step $t$. The difference between the maximum grid-block pressure $p_{i,max}^t$ and the minimum grid-block pressure $p_{i,min}^t$, for sample $i$ at time step $t$, is used to normalize the pressure absolute error.

To train the gas saturation and pressure buildup models we specify an initial learning rate of 0.0007 and 0.0006 for each model, respectively. The learning rate follows a 'staircase' reduction schedule. Following Wen et al. (2022), we train the gas saturation model for 100 epochs and the pressure buildup model for 140.

## 4. Results

In this section, we benchmark our U-DeepONet against U-FNO and Fourier-MIONet. The benchmarking criteria include: accuracy, training efficiency, and inference time. Note that the U-FNO was already benchmarked against the vanilla FNO and conv-FNO, which uses a CNN in place of the U-Net, and was shown to outperform both (Wen et al., 2022). Moreover, we perform our benchmarks on a NVIDIA Tesla V100. We implement our U-DeepONet using the open-source machine learning framework Pytorch (Paszke et al., 2019). Details of the U-DeepONet implementation are outlined in Table 2. The main difference between the saturation and the pressure buildup architectures is in the number of channels in the U-Net block. The code accompanying this manuscript is available upon request and the dataset was made publicly available courtesy of (Wen et al., 2022).

**Table 2:** U-DeepONet architectures. "Padding" denotes a padding operation. "Linear" denotes a linear transformation. "UNet2D" denotes a 2D U-Net. "ReLU" denotes the rectified linear unit. "C" denotes the batch size. **a.** is the gas saturation network architecture. **b.** is the pressure buildup network architecture.

|  | Part | Operation | Output shape |
|---|---|---|---|
| Input | – | – | (96, 200, 24, 12) |
| Padding | – | – | (C, 104, 208, 24, 12) |
| Trunk input | – | – | (24) |
| Branch input | – | – | (C, 96, 200, 12) |
| **a. Gas Saturation** | | | |
| Trunk net | FNN | Linear/Sin | (24, 64) |
| Branch net | Lifting (**P**) | Linear | (C, 104, 208, 64) |
|  | U-Net 1 | UNet2D/Sin | (C, 104, 208, 64) |
|  | U-Net 2 | UNet2D/Sin | (C, 104, 208, 64) |
|  | U-Net 3 | UNet2D/Sin | (C, 104, 208, 64) |
| Trunk-Branch product | Pointwise multiply | Hadamard product | (C, 104, 208, 24, 64) |
| Projection 1 (**G**) | Linear | Linear/ReLU | (C, 104, 208, 24, 64) |
| Projection 2 (**G**) | Linear | Linear/ReLU | (C, 104, 208, 24, 1) |
| De-padding | – | – | (C, 96, 200, 24) |
| **b. Pressure Buildup** | | | |
| Trunk net | FNN | Linear/Sin | (24, 96) |
| Branch net | Lifting (**P**) | Linear | (C, 104, 208, 96) |
|  | U-Net 1 | UNet2D/ReLU | (C, 104, 208, 96) |
|  | U-Net 2 | UNet2D/ReLU | (C, 104, 208, 96) |
|  | U-Net 3 | UNet2D/ReLU | (C, 104, 208, 96) |
| Trunk-Branch product | Pointwise multiply | Hadamard product | (C, 104, 208, 24, 96) |
| Projection 1 (**G**) | Linear | Linear/ReLU | (C, 104, 208, 24, 96) |
| Projection 2 (**G**) | Linear | Linear | (C, 104, 208, 24, 1) |
| De-padding | – | – | (C, 96, 200, 24) |

### 4.1. Gas Saturation

In this section, we train the U-DeepONet on the $CO_2$ sequestration dataset (Wen et al., 2022) to evaluate its performance and viability for learning gas saturation dynamics. The detailed architecture of the gas saturation U-DeepONet is outlined in Table 2a. We use a batch size of 4 to remain consistent with U-FNO (Wen et al., 2022) and Fourier-MIONet (Jiang et al., 2023) performance results. However, we remark here that using a larger batch size is made possible with our U-DeepONet due to its light memory footprint compared to the other two architectures, where the U-DeepONet requires only 4.6 GiB compared to 15.9 and 12.8 GiB for the U-FNO and Fourier-MIONet, respectively; see Table 3 for performance comparisons. This also gives the U-DeepONet an edge over other models as it is materialistically cheaper to train and deploy since many commercial GPUs have less than 8 GiB of memory.

Moreover, the U-DeepONet test set prediction error is lower than both the U-FNO and the Fourier-MIONet with an average test set MPE of only 1.58%. The testing set results, which contains 500 realizations, represent the predictability of the model on truly unseen data. In Fig. 5, four testing examples of U-DeepONet at two snapshots in time are shown. Using the MAE to benchmark, the U-DeepONet is about twice as accurate as the Fourier-MIONet and about 20% more accurate than the U-FNO. The U-DeepONet also has a higher $R^2$ score than the other two models.

In addition, U-DeepONet only requires about 108 seconds per epoch to train, i.e., it is about 18 times faster to train than the U-FNO on the same GPU, and about 5 times faster than the Fourier-MIONet, although the reported Fourier-MIONet performance is evaluated on a NVIDIA GeForce RTX 3090 GPU (Jiang et al., 2023) which has twice the CUDA cores. Not only is the U-DeepONet faster in training time, but it is also faster in testing time. For any given testing case, the U-DeepONet requires 0.016 seconds to predict the solution, while the U-FNO and the Fourier-MIONet require 0.018 and 0.041 seconds, respectively. Hence, these results show that the proposed U-DeepONet is more accurate and significantly more efficient in training and testing than both the U-FNO and Fourier-MIONet.

Our U-DeepONet is similar to the Fourier-MIONet in terms of flexibility in selecting the batch size and the number of time snapshots in the trunk input. This means that the accuracy and the computational efficiency of the U-DeepONet can potentially be further improved.

**Table 3**: Performance comparison among U-FNO, Fourier-MIONet, and U-DeepONet for gas saturation. Testing results are averaged over the entire test dataset.

| | Training | | | | | Testing | | | |
|---|---|---|---|---|---|---|---|---|---|
| Model | No. of parameters | GPU memory (GiB) | Training time / epoch (seconds) | Minimum epochs needed | Training time (hour) | $R^2$ | MPE (%) | MAE | Inference time (s) |
| U-FNO | 33,097,829 | 15.9 | 1912 | 100 | 53.1 | 0.981 | 1.61 | 0.0031 | 0.0182 |
| F-MIONet[*] | 3,685,325 | 12.8 | 730 | 77 ± 13 | 15.7 ± 2.7 | 0.982 ± 0.002 | - | 0.0050 | 0.041 |
| U-DeepONet | 1,803,369 | 4.6 | 108 | 100 | 3.0 | 0.994 | 1.58 | 0.0026 | 0.0156 |

[*]*The results of the Fourier-MIONet are reported from* (Jiang et al., 2023) *based on a NVIDIA GeForce RTX 3090 GPU since we do not have access to their code at the time of writing the manuscript. Also, the comparison is reported for the case of full batch size in time for consistency.*

### 4.2. Pressure Buildup

As mentioned earlier, a separate U-DeepONet is trained on the pressure buildup data. The architecture of this model is outlined in Table 2b. The performance evaluation metrics for pressure buildup outlined in Section 3.6 are reported in Table 4. As tabulated, the improvements in accuracy for the U-DeepONet over the U-FNO are not as apparent as in the gas saturation model, with the U-DeepONet having a higher $R^2$ score of 0.994 compared to 0.992 for the U-FNO and only 0.986 for Fourier-MIONet. The U-DeepONet also performs better than the U-FNO in terms of MAE (0.64 vs. 0.66), while the MAE of the Fourier-MIONet is not reported. In terms of the MRE (Eq. 22), the U-FNO performs slightly better than the U-DeepONet (0.0068 vs. 0.0072). In Section 5.1, we show that the MRE for the U-DeepOnet can be further dropped to 0.0069 if a batch size of 6 is used instead of 4.

It is not surprising that the U-DeepONet improvements in accuracy for the pressure buildup model were not as renowned as in the gas saturation case. This is because the U-FNO utilizes the Fourier transform which tends to perform well on smooth PDEs. This is particularly true in the case of the pressure equation which is diffusive by nature. Nonetheless, just as in the gas saturation model, the computational efficiency improvements are vast, with the U-DeepONet being about 15 times faster to train than the U-FNO and using almost a third of the GPU memory requirements. Furthermore, the U-DeepONet is faster than the U-FNO in inference. Fig. 6 shows four testing examples of predictions for pressure buildup. Our results for both saturation and pressure buildup models clearly indicate the advantages of the U-DeepONet over other models.

**Table 4:** Performance comparison among U-FNO, Fourier-MIONet, and U-DeepONet for pressure buildup. Testing results are averaged over the entire test dataset.

| Model | Training | | | | | Testing | | | |
|---|---|---|---|---|---|---|---|---|---|
| | No. of parameters | GPU memory (GiB) | Training time / epoch (s) | Minimum epochs needed | Training time (hour) | $R^2$ | MRE | MAE | Inference time (s) |
| U-FNO | 33,097,829 | 15.9 | 1912 | 140 | 74.4 | 0.992 | 0.0068 | 0.6585 | 0.0182 |
| F-MIONet[*] | - | - | - | - | - | 0.986 | - | - | - |
| U-DeepONet | 4,052,161 | 5.9 | 125 | 140 | 4.8 | 0.994 | 0.0072 | 0.6488 | 0.0154 |

[*]*The results of the Fourier-MIONet are reported in* (Jiang et al., 2023) *using a batch size of 8 for the time input. Other pertinent information in the table were not reported.*

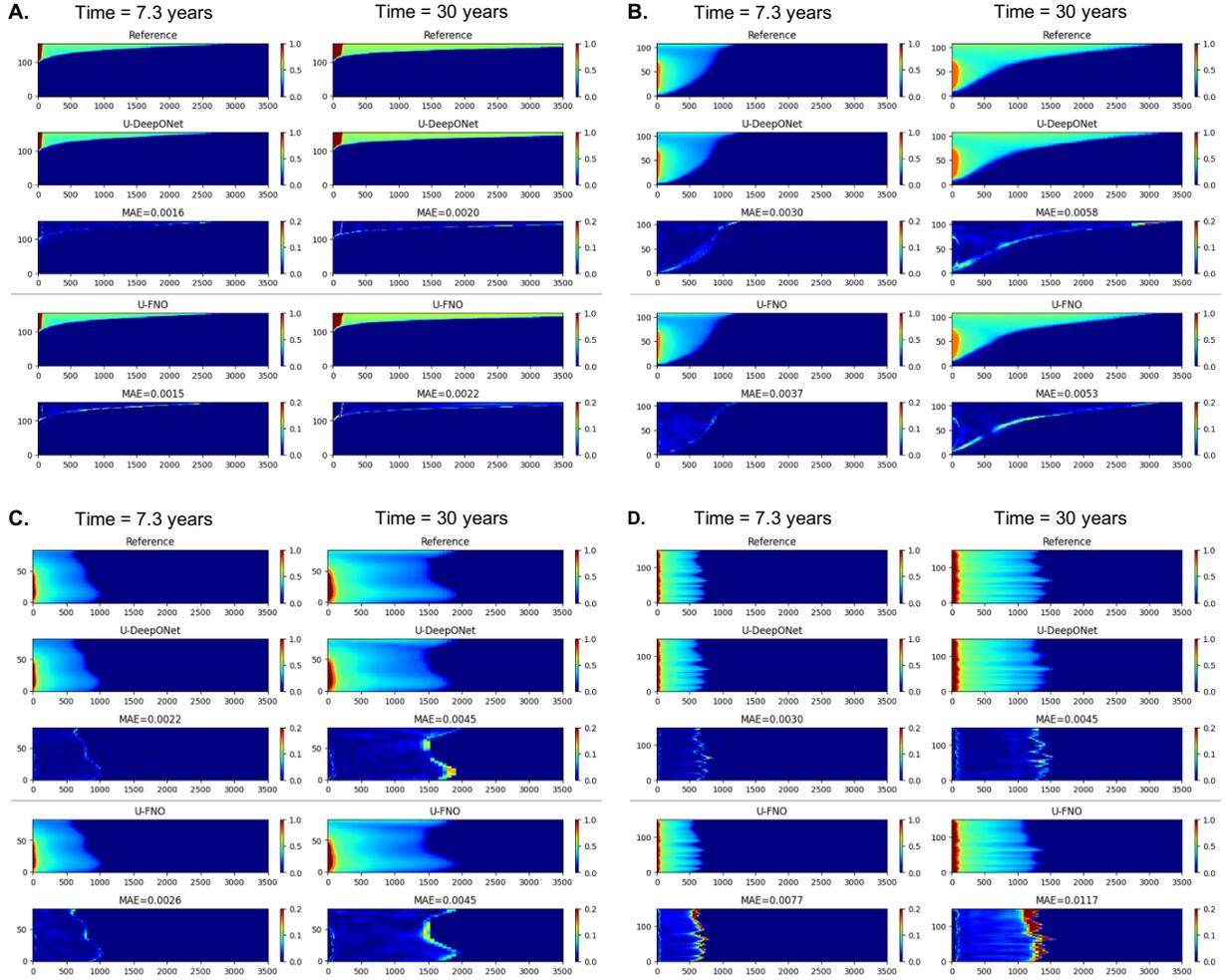

**Figure 5:** Visualizations of four testing examples of gas saturation. In each example, two time snapshots are shown at 7.3 and 30 years. The reference solution refers to the ground truth produced by the simulator. For each example we show the U-DeepONet and the U-FNO solutions along with associated absolute error map and the MAE.

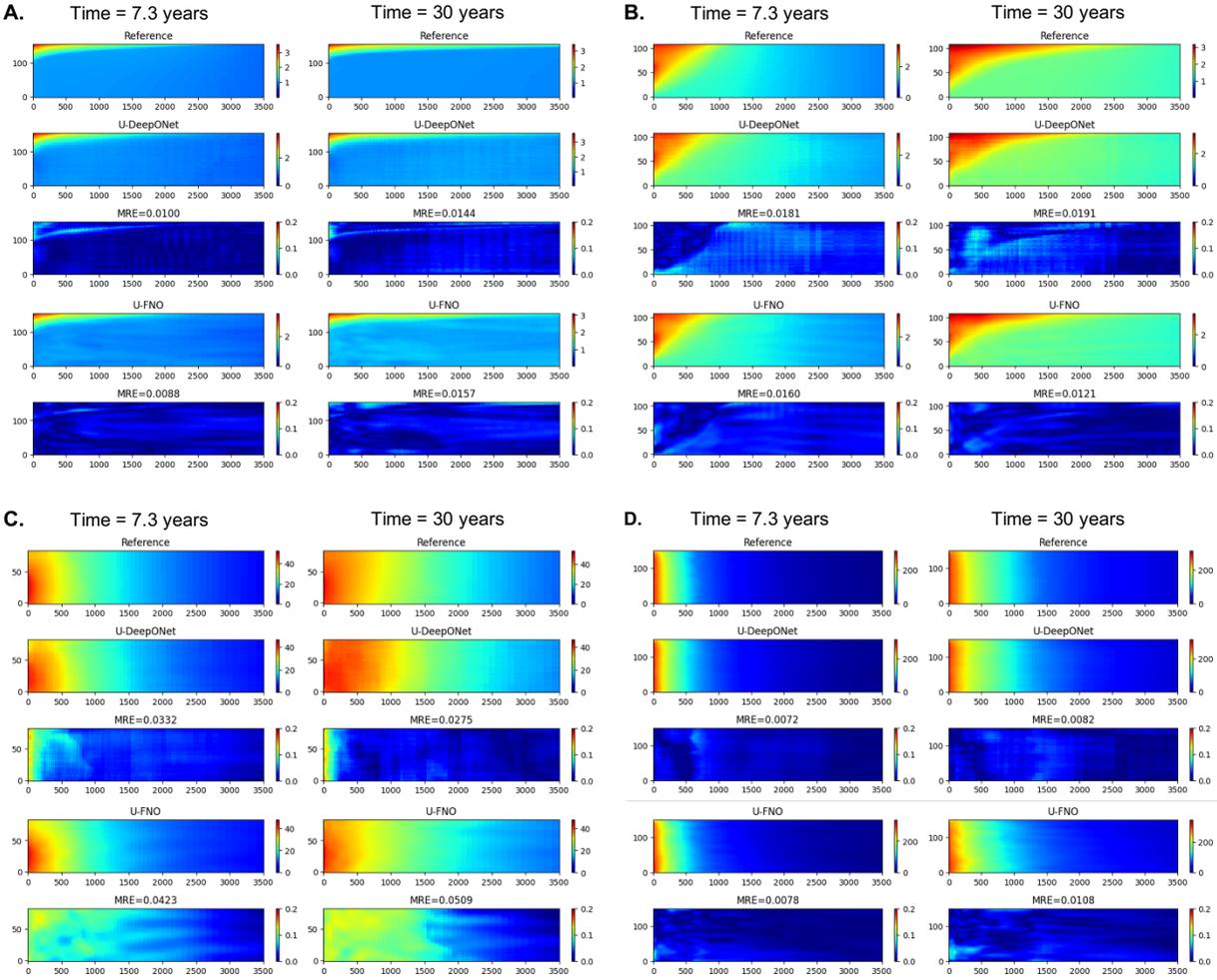

**Figure 6:** Visualizations of four testing examples of pressure buildup. In each example, two time snapshots are shown at 7.3 and 30 years. The reference solution refers to the ground truth produced by the simulator. For each example we show the U-DeepONet and the U-FNO solutions along with associated absolute error map and the MRE.

## 5. Discussion

### 5.1. Effect of batch size

So far, we have maintained the same batch size of 4 in our U-DeepONet for both the saturation and pressure buildup models to ensure a fair comparison with other architectures. However, the influence of batch size on model accuracy is complex and cannot be determined a priori. Here, we examine the effect of batch size on the accuracy of the U-DeepONet architecture. Fig. 7 illustrates how varying batch sizes affect the Mean Percentage Error (MPE) for the gas saturation model. Notably, increasing the batch size to 6 enhances the U-DeepONet's performance on both the gas saturation and pressure buildup datasets. Specifically, the MPE for the gas saturation model drops to 1.39%, and the Mean Relative Error (MRE) for the pressure buildup dataset decreases to 0.69%.

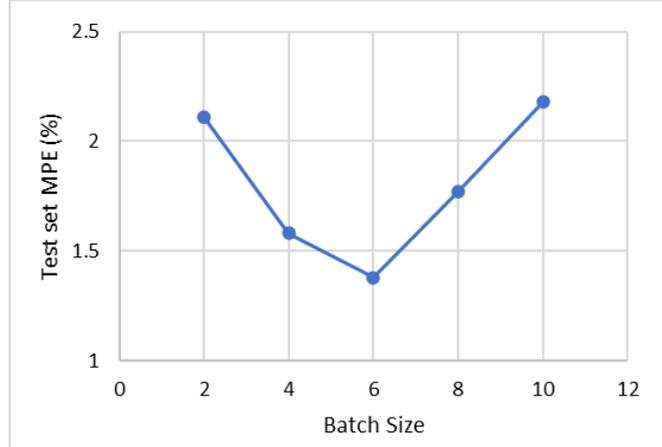

**Figure 7:** Effect of batch size on accuracy for the gas saturation model.

## 5.2. Data Efficiency

In Section 4, we demonstrated the superior training efficiency of the U-DeepONet compared to the U-FNO which resulted in considerably faster training times and more accurate predictions on the testing data set, especially for the gas saturation model. Building on this, we now present evidence that the U-DeepONet also excels in generalization, particularly in the context of the gas saturation dataset. This translates to a notable improvement in data utilization efficiency. The ability of U-DeepONet to effectively generalize with smaller datasets is significant, given the high computational cost of generating large datasets. This efficiency is not just a technical achievement; it has practical implications in reducing computational expenses and facilitating broader applications, especially in scenarios where data collection and acquisition is challenging.

To evaluate the generalizability of our U-DeepONet, we created nine subsets from the main training dataset through random subsampling. These subsets vary in size, ranging from 500 to the full set of 4500 realizations, to examine the model's performance across different data volumes. We report the MPE for the saturation test dataset and the MRE for the pressure buildup dataset in Fig. 8. This figure also includes results for testing sets using batch sizes of 4 and 6 to further understand the impact of batch size on model performance. To ensure the robustness of our findings and mitigate the influence of random subsampling, we repeated the training with different seeds to confirm the consistency of our results.

For the gas saturation dataset, even when trained with just 3000 realizations (two-thirds of the full dataset) and a batch size of 6, the U-DeepONet still surpasses the U-FNO, achieving a Mean Percentage Error (MPE) of 1.48%. This indicates that the U-DeepONet requires significantly less data to achieve superior testing set performance. Furthermore, in scenarios with limited training data, the U-DeepONet proves to be more reliable, maintaining a testing MPE of 3% when trained with only 500 realizations—about 10% of the full dataset. In contrast, the U-FNO needs over double that amount of data to reach a similar level of performance. Regarding the pressure buildup dataset, although the improvements in data utilization are less pronounced compared to the saturation case, the U-DeepONet still outperforms the U-FNO in small data regimes, ranging from 500 to 1500 realizations. The robust U-FNO performance on the pressure dataset is largely due to the Fourier transform which tends to perform well on smooth data.

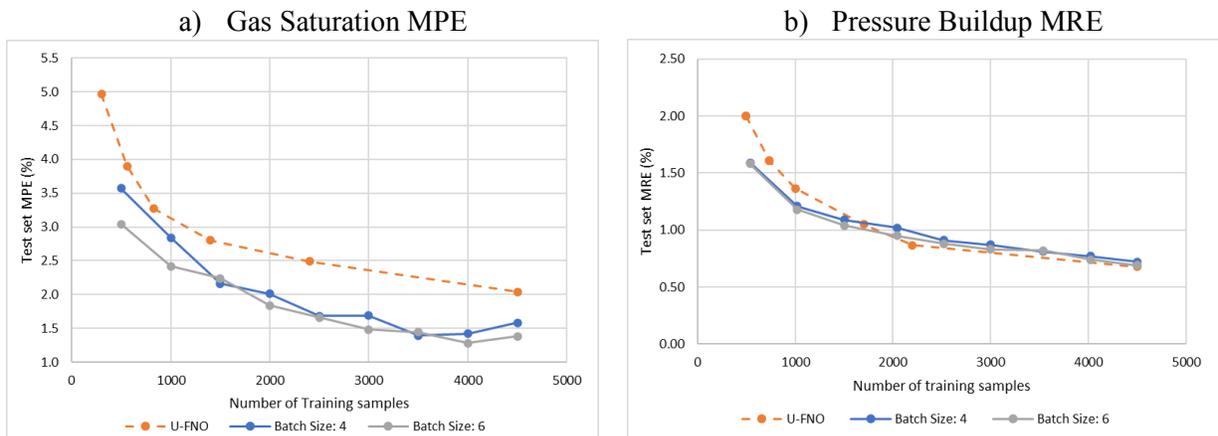

**Figure 8:** Generalization of the U-DeepONet with various batch sizes vs. U-FNO. a) Gas saturation testing dataset MPE vs. training dataset size. b) Pressure buildup testing dataset MRE vs. training dataset size.

## *5.3. Inference at unseen time steps*

One of the primary limitations of the U-FNO architecture is its grid-invariance in time, stemming directly from the use of a U-Net block to process all inputs, including the time variable. Unlike the U-FNO, both the Fourier-MIONet and our U-DeepONet circumvent this issue by processing the time variable through a separate network. In the U-DeepONet case, the time variable is passed to the trunk which allows us to make predictions at unseen time steps within the temporal training horizon. The advantage of this is twofold: it allows for further reduction in the size of the dataset required to train the neural operator, and it leads to a decrease in computational load when generating the dataset using a numerical solver. This reduction is achieved as fewer time steps are needed in numerical schemes, without compromising the accuracy of the machine learning model. Such feature is particularly valuable in applications where data collection is costly or challenging.

To demonstrate the U-DeepONet's capability to generalize to unseen time steps, we trained our models with only 13 time steps (~50% fewer time steps). Building upon the consistent performance of the U-DeepONet with a reduced dataset, as previously shown in sections 5.1 and 5.2, and to further demonstrate this capability, we augment the reduction in the time steps with a reduction in the overall size of the training dataset. Accordingly, we used 3500 realizations for the saturation model and 4000 for the pressure buildup model, while maintaining a batch size of 6 for both. To evaluate these models, we present the MPE in Figure 9a and the MRE in Figure 9b for the test datasets of the saturation and pressure buildup models, respectively, at each time step.

Training the U-DeepONet models with only ~50% of the time steps leads to a slight decrease in accuracy at the omitted time steps, as depicted by the empty squares in Figure 9. On the other hand, the performance of the U-DeepONet at time steps included in the training does not deteriorate (indicated by the solid squares in Figure 9). It is important to highlight that the first- and last-time steps should always be included in the training because data-driven neural operator learning methods cannot usually extrapolate in time.

From Figure 9, it's evident that the U-DeepONet's performance (gray curves with squares) using the hyperparameters discussed in this paper falls short at unseen time steps. To address this, we focused on the temporal interpolation task by refining the trunk network. The initial trunk network featured 10 layers and 64 neurons per layer for the saturation model and 96 neurons per layer for the pressure model. The updated design comprises 16 layers and 14 neurons per layer for the first 15 layers, while the last layer in each model maintains 64 and 96 neurons for the gas and pressure models, respectively, to enable the dot product with the branch. The results of this modification are shown in Figure 9 (depicted by the orange curve with diamonds). We can see from figures 9a and 9b that the U-DeepONet regains its stellar performance at unseen time steps due to the fine-tuning of the trunk network.

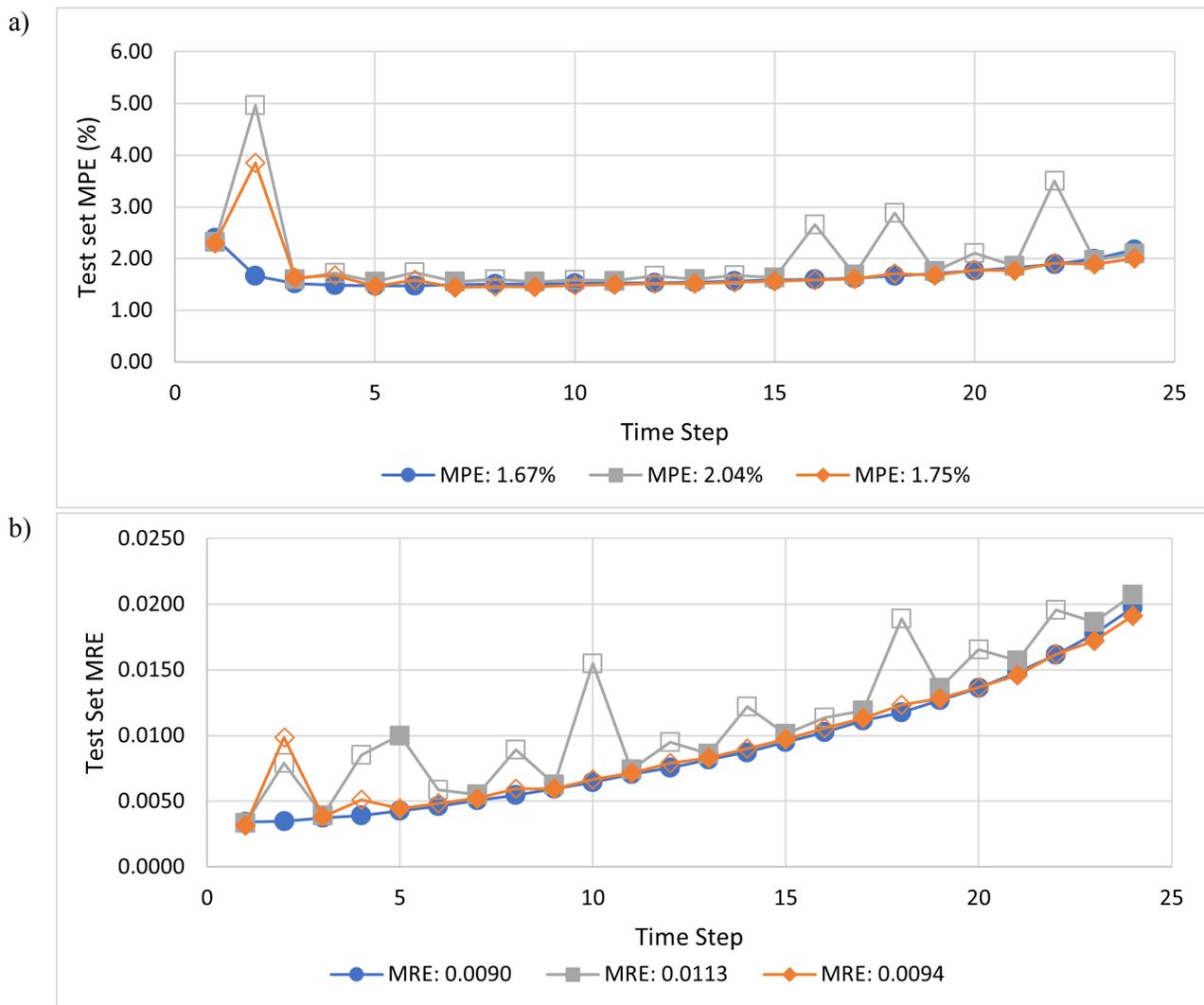

**Figure 9:** U-DeepONet performance per time step. Blue curve is the reference solution with all the available time steps. Gray curve is the solution with only 13 timesteps during training and without fine-tuning the trunk network. Orange curve is the solution with only 13 timesteps during training and a fine-tuned trunk network. Solid shapes mean the time step was included in the training, and empty shapes mean that the time step was omitted from the training set. a) per step MPE of the saturation test set. b) per step MRE of the pressure buildup test set.

## 6. Conclusions

This paper presents a novel U-Net enhanced deep operator network, termed U-DeepONet. In designing the U-DeepONet, the best features from both the U-FNO and the Fourier-MIONet were fused into an architecture that outperforms the other two architectures in training and testing performance for multiphase flow and transport problems in porous media. We evaluate the novel U-DeepONet using the open-source $CO_2$ sequestration dataset (Wen et al., 2022) and compare performance to that of the U-FNO and the Fourier-MIONet. Results show that the U-DeepONet is advantageous in performance, predictive accuracy, training efficiency, and data utilization efficiency. Our U-DeepONet is more than 18 times faster in training than the U-FNO and more than 5 times faster than the Fourier-MIONet, while being more accurate than both models. We also show that the U-DeepONet has a much smaller GPU memory footprint compared to the other operator learning algorithms and is faster in inference. The U-DeepONet is data efficient and better at generalization as it can be trained with less data while maintaining accuracy. Moreover, we show that the U-DeepONet performance at unseen time steps is robust without any changes to the architecture. Overall, we show that the U-DeepONet is a better framework for neural operator learning compared to other state-of-the-art frameworks, and that the U-DeepONet is easier to work with due to the smaller number of hyperparameters. Finally, the U-DeepONet architecture can be seen as an alternative to the multiple-input DeepONet (MIONet), as we have shown that it can learn multiple operators simultaneously.

## Data Availability

The raw/processed data required to reproduce these findings is available at https://drive.google.com/drive/folders/1fZQfMn_vsjKUXAfRV0q_gswtl8JEkVGo?usp=sharing courtesy of [43].

## CRediT authorship contribution statement

**Waleed Diab**: Conceptualization, Methodology, Software, Formal analysis, Investigation, Validation, Visualization, Writing – original draft, Writing – review & editing. **Mohammed Al Kobaisi**: Supervision, Conceptualization, Methodology, Formal analysis, Validation, Writing – original draft, Writing – review & editing.

## Declaration of competing interest

The authors declare that they have no known competing financial interests or personal relationships that could have appeared to influence the work reported in this paper.

## Acknowledgement

The authors wish to acknowledge Khalifa University's high-performance computing facilities for providing the computational resources.

## 7. References

Ajayi, T., Gomes, J. S., & Bera, A. (2019). A review of CO2 storage in geological formations emphasizing modeling, monitoring and capacity estimation approaches. In *Petroleum Science* (Vol. 16, Issue 5, pp. 1028–1063). China University of Petroleum Beijing. https://doi.org/10.1007/s12182-019-0340-8

Alpak, F. O., Vamaraju, J., Jennings, J. W., Pawar, S., Devarakota, P., & Hohl, D. (2023). Augmenting Deep Residual Surrogates with Fourier Neural Operators for Rapid Two-Phase Flow and Transport Simulations. *2023 SPE Journal*, *1*. https://doi.org/10.2118/217441-PA/3202043/spe-217441-pa.pdf

Anyosa, S., Bunting, S., Eidsvik, J., Romdhane, A., & Bergmo, P. (2021). Assessing the value of seismic monitoring of CO2 storage using simulations and statistical analysis. *International Journal of Greenhouse Gas Control*, *105*. https://doi.org/10.1016/j.ijggc.2020.103219

Bachu, S. (2008). CO2 storage in geological media: Role, means, status and barriers to deployment. In *Progress in Energy and Combustion Science* (Vol. 34, Issue 2, pp. 254–273). https://doi.org/10.1016/j.pecs.2007.10.001

Benson, S. M., & Cole, D. R. (2008). CO2 sequestration in deep sedimentary formations. *Elements*, *4*(5), 325–331. https://doi.org/10.2113/gselements.4.5.325

Cao, C., Liao, J., Hou, Z., Wang, G., Feng, W., & Fang, Y. (2020). Parametric uncertainty analysis for CO2 sequestration based on distance correlation and support vector regression. *Journal of Natural Gas Science and Engineering*, *77*. https://doi.org/10.1016/j.jngse.2020.103237

Cardoso, M. A., Durlofsky, L. J., & Sarma, P. (2009). Development and application of reduced-order modeling procedures for subsurface flow simulation. *International Journal for Numerical Methods in Engineering*, *77*(9), 1322–1350. https://doi.org/10.1002/nme.2453

Espeholt, L., Agrawal, S., Sønderby, C., Kumar, M., Heek, J., Bromberg, C., Gazen, C., Carver, R., Andrychowicz, M., Hickey, J., Bell, A., & Kalchbrenner, N. (2022). Deep learning for twelve hour precipitation forecasts. *Nature Communications*, *13*(1). https://doi.org/10.1038/s41467-022-32483-x

Falola, Y., Misra, S., & Nunez, A. C. (2023, October 2). Rapid High-Fidelity Forecasting for Geological Carbon Storage Using Neural Operator and Transfer Learning. *Day 1 Mon, October 02, 2023*. https://doi.org/10.2118/216135-MS

Fawad, M., & Mondol, N. H. (2021). Monitoring geological storage of CO2: a new approach. *Scientific Reports*, *11*(1). https://doi.org/10.1038/s41598-021-85346-8

Flemisch, B., Nordbotten, J. M., Fernø, M., Juanes, R., Both, J. W., Class, H., Delshad, M., Doster, F., Ennis-King, J., Franc, J., Geiger, S., Gläser, D., Green, C., Gunning, J., Hajibeygi, H., Jackson, S. J., Jammoul, M., Karra, S., Li, J., … Zhang, Z. (2023). The FluidFlower Validation Benchmark Study for the Storage of CO 2. *Transport in Porous Media*. https://doi.org/10.1007/s11242-023-01977-7

Gan, M., Nguyen, M. C., Zhang, L., Wei, N., Li, J., Lei, H., Wang, Y., Li, X., & Stauffer, P. H. (2021). Impact of reservoir parameters and wellbore permeability uncertainties on CO2 and brine leakage potential at the Shenhua CO2 Storage Site, China. *International Journal of Greenhouse Gas Control*, *111*. https://doi.org/10.1016/j.ijggc.2021.103443

Goswami, S., Bora, A., Yu, Y., & Karniadakis, G. E. (2022). *Physics-Informed Deep Neural Operator Networks*. 1–34. http://arxiv.org/abs/2207.05748

Jeong, H., Srinivasan, S., & Bryant, S. (2013). Uncertainty quantification of CO2 plume migration using static connectivity of geologic features. *Energy Procedia*, *37*, 3771–3779. https://doi.org/10.1016/j.egypro.2013.06.273

Jiang, Z., Zhu, M., Li, D., Li, Q., Yuan, Y. O., & Lu, L. (2023). *Fourier-MIONet: Fourier-enhanced multiple-input neural operators for multiphase modeling of geological carbon sequestration*. arXiv:2303.04778v1

Jin, P., Meng, S., & Lu, L. (2022). MIONet: Learning Multiple-Input Operators via Tensor Product. *Https://Doi.Org/10.1137/22M1477751*, *44*(6), A3490–A3514. https://doi.org/10.1137/22M1477751

Ju, X., Hamon, F. P., Wen, G., Kanfar, R., Araya-Polo, M., & Tchelepi, H. A. (2023). *Learning CO2 plume migration in faulted reservoirs with Graph Neural Networks*. http://arxiv.org/abs/2306.09648

Kissas, G., Seidman, J., Guilhoto, L. F., Preciado, V. M., Pappas, G. J., & Perdikaris, P. (2022). Learning Operators with Coupled Attention. *Journal of Machine Learning Research*, *23*, 1–63. http://arxiv.org/abs/2201.01032

Lengler, U., De Lucia, M., & Kühn, M. (2010). The impact of heterogeneity on the distribution of CO2: Numerical simulation of CO2 storage at Ketzin. *International Journal of Greenhouse Gas Control*, *4*(6), 1016–1025. https://doi.org/10.1016/j.ijggc.2010.07.004

Li, Z., Kovachki, N., Azizzadenesheli, K., Liu, B., Bhattacharya, K., Stuart, A., & Anand. (2021). *Fourier neural operator for parametric partial differential equations*. arXiv:2010.08895v3

Lichtner, P., Karra, S., Hammond, G., Lu, C., Bisht, G., Kumar, J., Mills, R., & Andre, B. (n.d.). *PFLOTRAN User Manual: A Massively Parallel Reactive Flow and Transport Model for Describing Surface and Subsurface Processes*.

Lu, L., Jin, P., Pang, G., Zhang, Z., & Karniadakis, G. E. (2021a). DeepONet: Learning nonlinear operators via DeepONet based on the universal approximation theorem of operators. *Nature Machine Intelligence*, *3*(3), 218–229. https://doi.org/10.1038/s42256-021-00302-5

Lu, L., Jin, P., Pang, G., Zhang, Z., & Karniadakis, G. E. (2021b). Learning nonlinear operators via DeepONet based on the universal approximation theorem of operators. *Nature Machine Intelligence*, *3*(3), 218–229. https://doi.org/10.1038/s42256-021-00302-5

Lu, L., Meng, X., Cai, S., Mao, Z., Goswami, S., Zhang, Z., & Karniadakis, G. E. (2022). A comprehensive and fair comparison of two neural operators (with practical extensions) based on FAIR data. *Computer Methods in Applied Mechanics and Engineering*, *393*, 1–35. https://doi.org/10.1016/j.cma.2022.114778

Lyu, Y., Zhao, X., Gong, Z., Kang, X., & Yao, W. (2023). Multi-fidelity prediction of fluid flow based on transfer learning using Fourier neural operator. *Physics of Fluids*, *35*(7). https://doi.org/10.1063/5.0155555

Mahjour, S. K., & Faroughi, S. A. (2023). Selecting representative geological realizations to model subsurface CO2 storage under uncertainty. *International Journal of Greenhouse Gas Control*, *127*. https://doi.org/10.1016/j.ijggc.2023.103920

Nordbotten, J. M., Flemisch, B., Gasda, S. E., Nilsen, H. M., Fan, Y., Pickup, G. E., Wiese, B., Celia, M. A., Dahle, H. K., Eigestad, G. T., & Pruess, K. (2012). Uncertainties in practical simulation of CO2 storage. *International Journal of Greenhouse Gas Control*, *9*, 234–242. https://doi.org/10.1016/j.ijggc.2012.03.007

Paszke, A., Gross, S., Massa, F., Lerer, A., Bradbury, J., Chanan, G., Killeen, T., Lin, Z., Gimelshein, N., Antiga, L., Desmaison, A., Köpf, A., Yang, E., DeVito, Z., Raison, M., Tejani, A., Chilamkurthy, S., Steiner, B., Fang, L., … Chintala, S. (2019). PyTorch: An imperative style, high-performance deep learning library. *Advances in Neural Information Processing Systems*, *32*(NeurIPS).

Pruess, K., & García, J. (2002). Multiphase flow dynamics during CO2 disposal into saline aquifers. *Environmental Geology*, *42*(2–3), 282–295. https://doi.org/10.1007/s00254-001-0498-3

Saadatpoor, E., Bryant, S. L., & Sepehrnoori, K. (2010). New trapping mechanism in carbon sequestration. *Transport in Porous Media*, *82*(1), 3–17. https://doi.org/10.1007/s11242-009-9446-6

Schlumberger. (2014). *ECLIPSE reservoir simulation software reference manual*.

Stepien, M., Ferreira, C. A. S., Hosseinzadehsadati, S., Kadeethum, T., & Nick, H. M. (2023). Continuous conditional generative adversarial networks for data-driven modelling of geologic CO2 storage and plume evolution. *Gas Science and Engineering*, *115*. https://doi.org/10.1016/j.jgsce.2023.204982

Strandli, C. W., Mehnert, E., & Benson, S. M. (2014). CO2 plume tracking and history matching using multilevel pressure monitoring at the Illinois basin - Decatur project. *Energy Procedia*, *63*, 4473–4484. https://doi.org/10.1016/j.egypro.2014.11.483

Tang, M., Ju, X., & Durlofsky, L. J. (2022). Deep-learning-based coupled flow-geomechanics surrogate model for CO2 sequestration. *International Journal of Greenhouse Gas Control*, *118*. https://doi.org/10.1016/j.ijggc.2022.103692

Tang, M., Liu, Y., & Durlofsky, L. J. (2020). A deep-learning-based surrogate model for data assimilation in dynamic subsurface flow problems. *Journal of Computational Physics*, *413*. https://doi.org/10.1016/j.jcp.2020.109456

Tariq, Z., Ali, M., Yan, B., Sun, S., Khan, M., Yekeen, N., & Hoteit, H. (2023). Data-Driven Machine Learning Modeling of Mineral/CO2/Brine Wettability Prediction: Implications for CO2 Geo-Storage. *SPE Middle East Oil and Gas Show and Conference, MEOS, Proceedings*. https://doi.org/10.2118/213346-MS

Tariq, Z., Yan, B., & Sun, S. (2023). Predicting Trapping Indices in CO2 Sequestration- A Data-Driven Machine Learning Approach for Coupled Chemo-Hydro-Mechanical Models in Deep Saline Aquifers. *ARMA US Rock Mechanics/Geomechanics Symposium*. https://doi.org/https://doi.org/10.56952/ARMA-2023-0757

Tripura, T., & Chakraborty, S. (2022). *Wavelet neural operator: a neural operator for parametric partial differential equations*. http://arxiv.org/abs/2205.02191

Wang, S., & Perdikaris, P. (2021). Long-time integration of parametric evolution equations with physics-informed DeepONets. *Journal of Computational Physics*, *475*, 111855. https://doi.org/10.1016/j.jcp.2022.111855

Wang, S., Wang, H., & Perdikaris, P. (2021). Learning the solution operator of parametric partial differential equations with physics-informed DeepONets. *Science Advances*, *7*(40). https://doi.org/10.1126/sciadv.abi8605

Wang, S., Wang, H., & Perdikaris, P. (2022). Improved Architectures and Training Algorithms for Deep Operator Networks. *Journal of Scientific Computing*, *92*(2), 1–42. https://doi.org/10.1007/s10915-022-01881-0

Wen, G., Li, Z., Azizzadenesheli, K., Anandkumar, A., & Benson, S. M. (2022). U-FNO—An enhanced Fourier neural operator-based deep-learning model for multiphase flow. *Advances in Water Resources*, *163*. https://doi.org/10.1016/j.advwatres.2022.104180

Wen, G., Li, Z., Long, Q., Azizzadenesheli, K., Anandkumar, A., & Benson, S. M. (2023). Real-time high-resolution CO2 geological storage prediction using nested Fourier neural operators. *Energy and Environmental Science*, *16*(4), 1732–1741. https://doi.org/10.1039/d2ee04204e

Xiao, C., Zhang, S., Ma, X., Zhou, T., Hou, T., & Chen, F. (2023). Deep-learning-generalized data-space inversion and uncertainty quantification framework for accelerating geological CO2 plume migration monitoring. *Geoenergy Science and Engineering*, *224*. https://doi.org/10.1016/j.geoen.2023.211627

Yan, B., Chen, B., Robert Harp, D., Jia, W., & Pawar, R. J. (2022). A robust deep learning workflow to predict multiphase flow behavior during geological CO2 sequestration injection and Post-Injection periods. *Journal of Hydrology*, *607*. https://doi.org/10.1016/j.jhydrol.2022.127542

Yin, Z., Siahkoohi, A., Louboutin, M., & Herrmann, F. J. (2022). Learned coupled inversion for carbon sequestration monitoring and forecasting with Fourier neural operators. *SEG Technical Program Expanded Abstracts*, *2022-August*, 467–472. https://doi.org/10.1190/image2022-3722848.1

Zhang, K., Wu, Y.-S., & Pruess, K. (2008). *User's Guide for TOUGH2-MP-A Massively Parallel Version of the TOUGH2 Code*.

Zhang, K., Zuo, Y., Zhao, H., Ma, X., Gu, J., Wang, J., Yang, Y., Yao, C., & Yao, J. (2022). Fourier Neural Operator for Solving Subsurface Oil/Water Two-Phase Flow Partial Differential Equation. *SPE Journal*, *27*(3), 1815–1830. https://doi.org/10.2118/209223-PA

Zhao, M., Wang, Y., Gerritsma, M., & Hajibeygi, H. (2023). Efficient simulation of CO2 migration dynamics in deep saline aquifers using a multi-task deep learning technique with consistency. *Advances in Water Resources*, *178*. https://doi.org/10.1016/j.advwatres.2023.104494